\documentstyle[prd,preprint,aps]{revtex}
\newcommand{\es}{\mbox{\hspace{3mm}}}
\begin{document}
\tightenlines
\parskip 0.3cm
\begin{titlepage}

\begin{centering}
{\large \bf $\tau$ magnetic moment in a $\gamma \gamma$ collider}\\
\vspace{.9cm}
{\bf L.Tabares and O.A.Sampayo}\\
\vspace{.05in}
{\it  Departamento de F\'{\i}sica,
Universidad Nacional de Mar del Plata \\
Funes 3350, (7600) Mar del Plata, Argentina} \\ \vspace{.4cm}
\vspace{.08in}
{ \bf Abstract}
\\
\bigskip
\end{centering}
{\small We analyze different observables to study the magnetic
dipole moment of the tau pairs produced by photon linear
colliders. We use the circular polarized photon beam and study
distributions of polarized $\tau$ final pairs to define sensibly
asymmetries to magnetic dipole moment.}


\vspace{0.2in}
\vfill

\end{titlepage}
\vfill\eject

\section{\bf Introduction}

The bounds on the anomalous magnetic moment of the $\tau$ lepton,
$a_{\tau}$ are much weaker than the ones for electron and muon.
Due to the $\tau$ lepton's short lifetime of $(291.0\pm1.5)\times
10^{-15}$ seg, its anomalous magnetic moment can not be measured
by a spin precession method and no direct measurement of
$a_{\tau}$ exists so far. Because of the impossibility of
measuring $a_{\tau}$ by spin precession method, the present bounds
have been obtained by analysis of collision experiments. In that
sense, interesting articles, which obtain bounds on $a_{\tau}$,
have been published recently. The OPAL Collaboration \cite{opal}
uses a reaction proposed for A. Mendez and A. Grifolds
\cite{mendez} some years ago. They obtained limits on $a_{\tau}$
from the non-observation of anomalous $\tau \bar{\tau} \gamma$
production at LEP. In other article, G.A.Gonzales-Sprinberg,
A.Santamaria and J.Vidal using LEP1, SLD, and LEP2 data, for tau
lepton production, and data from CDF, D0 and LEP2, for W-decays
into tau lepton, established model independent limits on
no-standard electromagnetic and weak magnetic moments of the tau
lepton \cite{gabriel}.  In this case the obtained electromagnetic
bound is $\mid a_{\tau}\mid < 0.009$ at 95\% C.L., which is the
best bound obtained today but still far away of the theoretical
precision for $a_{\tau}$:$(1177.3 \pm 0.3)\times 10^{-6}$. This
weakness of the $a_{\tau}$ bounds is unfortunate since large
deviations from the S.M. values are expected for the $\tau$
lepton. In particular, in composite models one would expect larger
effects for the tau lepton than for the rest of the leptons.
 In this article
we are interested in studying the capability of an $\gamma \gamma$
collider to improve the existing bounds on $a_{\tau}$. In this
respect we study the $\gamma \gamma \rightarrow \tau \bar\tau$
process which was studied in the past for different authors
\cite{gg}. It is an extremely clean process because it has not
interference with weak interactions, being a purely QED process.
Moreover, the high center of mass energies proposed for this
collider makes of it an adequate laboratory to see anomalous
magnetic moment effects which grow with the energy. In this work,
we analyze a different set of observables than the ones previously
studied by other authors \cite{gg}. In particular, we analyze the
angular distribution for the universality ratio and asymmetries in
the production of polarized $\tau$ when the initial photons are
polarized too. As far as we know this observable set has not been
analyzed yet.

 Following Ref \cite{masso}, in order to analyze tau magnetic moments, we will
use an effective lagrangian description. Thus, in section 2, we
describe the effective lagrangian formalism. In section 3, we
summary the helicity formalism, which is used for calculating the
cross-section. In the same section we present the different used
observable. Finally, in section 4 we give our conclusions.

\section{\bf The effective lagrangian approach}

In the last few years, effective lagrangians have been used as an
adequate framework to study low energies effects of physics beyond
the standard model (SM). Since the SM gives a very good
description of all physics at energies available at present
accelerators, then one expects that any deviation of the SM can be
parametrized by an effective lagrangian built with the field and
symmetries of the SM. In this conditions, the effective lagrangian
contains a renormalizable piece, the SM theory and
non-renormalizable operators of dimension higher than 4 which are
suppresed by inverse powers of the high energy physics scale,
$\Lambda$. The leading non-standard effects will come from the
operators with the lowest dimension. Those are dimension six
operators. In particular, there are only two six-dimension operators which contribute to the anomalous magnetic
moments \cite{buchmuller}:

\begin{eqnarray}
{\cal O}_B&=&\frac{g'}{2 \Lambda^2} \bar L_L \phi \sigma_{\mu\nu}
\tau_R B^{\mu\nu} \\ {\cal O}_W&=&\frac{g}{2 \Lambda^2} \bar L_L
\overrightarrow{\tau} \phi \sigma_{\mu\nu} \tau_R
\overrightarrow{W}^{\mu\nu} \nonumber
\end{eqnarray}

where $L_L=(\nu_L,\tau_L)$ is the tau leptonic doublet and $\phi$
is the Higgs doublet. $B^{\mu\nu}$ and $W^{\mu\nu}$ are
the $U(1)_Y$ and $SU(2)_L$ field strength tensor, and $g'$ and $g$
are the corresponding gauge couplings. Thus, we write our
effective Lagrangian as

\begin{eqnarray}
{\cal L}_{eff} = {\cal L}_{SM} + \alpha_B {\cal O}_B + \alpha_W
{\cal O}_W + h.c.
\end{eqnarray}

As we are not interested in studying $CP$ violation effects, then we
will take the coupling $\alpha_B$ and $\alpha_W$ real. Moreover,
we will consider them as free parameters without any further
assumption.

The interaction lagrangian can be written in term of the physical
fields $A_{\mu}$, $Z_{\mu}$ and $W^{\pm}_{\mu}$. In our particular
case, we are only interested in the effective electromagnetic
interaction, since we are studying a process which only involves
electromagnetic interactions. In this conditions the relevant
lagrangians is

\begin{eqnarray}
{\cal L}_{eff}={\cal L}_{SM}+a_{\tau} \frac{e}{4 m_{\tau}} \bar
\tau \sigma_{\mu\nu} \tau F^{\mu\nu} + \cdots
\end{eqnarray}

where the dots represent non-relevant pieces of the lagrangian and
$F_{\mu\nu}$ is the electromagnetic field strength tensor. We have
expressed the coupling in function of $a_{\tau}$ because it is
directly related to the experimental measurement and theoretical
calculations. The $a_{\tau}$ constant can be expressed in terms of
$\alpha_B$, $\alpha_W$ and $\Lambda$ as follows

\begin{equation}
a_{\tau}=\frac{\sqrt{2} v m_{\tau}}{\Lambda^2} (\alpha_B-\alpha_W)
\end{equation}

\section{\bf The $\gamma \gamma \rightarrow \tau \bar \tau$ process}

In this section we study the $\gamma \gamma \rightarrow \tau \bar
\tau$ process which only involves electromagnetic interactions
plus additional magnetic moment couplings given by eq(3).

This process is of interest for a number of reasons. Increased
cross section for high energy and the absence of weak
contributions are further complementary features of the two-photon
process. In addition, very hard photons at high luminosity may be
produced in Compton backscattering of laser light of high energy
$e^+ e^-$ beams.

The corresponding Feynman diagrams are shown in Fig.1 and the
corresponding amplitude for the process can be written as:

\begin{equation}
{\cal M}=-i e^2 [P_{\tau}(p_1-k_1) T_1+P_{\tau}(p_1-k_2) T_2]
\end{equation}

where $P_{\tau}(k)=1/(k^2-m_{\tau}^2)$ and

\begin{eqnarray}
T_1&=&\bar u(p_1)(\gamma_{\mu}-i \frac{a_{\tau}}{2 m_{\tau}}
\sigma_{\mu\delta} k_1^{\delta})(p\llap{/}_1-k\llap{/}_1)
\nonumber \\ &&(\gamma_{\nu}-i \frac{a_{\tau}}{2 m_{\tau}}
\sigma_{\nu\rho} k_2^{\rho})v(p_2) \epsilon^{\mu}_1
\epsilon^{\nu}_2 \nonumber \\ T_2&=&\bar u(p_1)(\gamma_{\nu}-i
\frac{a_{\tau}}{2 m_{\tau}} \sigma_{\nu\rho}
k_2^{\rho})(p\llap{/}_1-k\llap{/}_2) \nonumber \\
 &&(\gamma_{\mu}-i
\frac{a_{\tau}}{2 m_{\tau}} \sigma_{\mu\delta} k_1^{\delta})v(p_2)
\epsilon^{\mu}_1  \epsilon^{\nu}_2
\end{eqnarray}

Bounds on the anomalous coupling $a_{\tau}$ can be obtained from a
test of universality in $\gamma \gamma$-colliders by assuming that
only the tau lepton has anomalous magnetic moment (muon and
electron magnetic moments have been measured quite precisely
\cite{caso}).

In order to compare with the experimental data, we define the {\it
angular dependent universality ratio} for unpolarized initial
photon and final lepton:

\begin{eqnarray}
R(\theta)=\frac{d \sigma/d \Omega |_{(\gamma \gamma \rightarrow
\tau \bar{\tau})}}{d \sigma/d \Omega |_{(\gamma \gamma \rightarrow
\mu \bar{\mu})}}
\end{eqnarray}

 This ratio is a function
of the dispersion angle and the tau lepton anomalous magnetic moment,
$a_{\tau}$. We have considered final lepton as massless.
In this conditions the above ratio can be written as

\begin{equation}
R(\theta) \simeq 1 + \delta R(\theta,a_{\tau})
\end{equation}

where

\begin{equation}
\delta R(\theta,a_{\tau})=4 c^2_{\tau} s
\frac{\sin^2(\theta)}{1+\cos^2(\theta)} \left[4+c_{\tau}^2 s
\sin^2(\theta)\right]
\end{equation}

For simplicity in the calculation we have used the related
constant $c_{\tau}=a_{\tau}/(4 m_{\tau})$. We illustrate the
behaviour of $\delta R$ as a function of $a_{\tau}$ and $\theta$,
showing in Fig.2 the contour plot for different values of $\delta
R$ in the ($\cos\theta$,$a_{\tau}$) plane for $\sqrt{s}=200$ GeV.
As it is shown in this figure the maximal sensibility of this
observable is reached for $\theta$ closed to $\pi / 2$.

Another way to present the same results is shown in Fig.3, where
we plot $R$ as a function of $\cos\theta$ for different values of
$a_{\tau}$, for $\sqrt{s}=200$ GeV. If the tau lepton is a
sequential lepton with the same properties of the electron and
muon, then we expect an $R=1$ value. In the other hand, if effects
of new physics were relevant for the tau lepton, then we would expect a
different value of $R$. For example, we consider a measurement of
$R$ compatible with one (the SM value) within a accuracy of
$10\%$, then we could obtain an $a_{\tau}$ bound of
$a_{\tau}\lesssim 0.003$ for $\sqrt{s}=200$ GeV.

The Photon Linear Collider may be the best alternative to the
electron-positron collider. In this collider we have the
opportunity to control the initial photon polarization by the
inverse Compton scattering of the polarized laser by the
electron-positron beam at NLC \cite{ginzburg,hori}. Adjusting the
laser polarization, we can get highly polarized photon. We discuss
only the case of circular polarized beam. In this conditions, we
study the process $\gamma \gamma \rightarrow \tau \bar{\tau}$ for
the polarized tau lepton pair production from circular polarized
photon collision ($\gamma_{L,R}\gamma_{L,R}\rightarrow \tau_{L,R}
\bar{\tau}_{L,R}$) in the center-mass (CM) frame at tree level
in perturbation theory. Due to the high CM energies involved, we can
consider the final lepton as massless. The polarization of the
final tau (anti-tau) can be studied by their decay products. They
are referred to as the spin analyzer of the tau lepton. Effectively,
the spin polarization of the produced tau is reflected in the
distorted distribution of the decay products. Therefore, the
$\tau$-polarization can be determined from a measurement of the
spectrum of the final charged particle in the following decay
channels: $\tau^- \rightarrow \nu_{\tau} \pi^-$,$\nu_{\tau} \rho^-
$,$\nu_{\tau} a_1^-$,$\nu_{\tau} e^- \bar{\nu}_e$,$\nu_{\tau}
\mu^- \bar{\nu}_{\mu}$.

In order to simplify the calculation of the cross-section when the initial photons
and the final tau pairs are polarized we use the Helicity
Amplitude Method (HAM). In this section we summarize the principal features of
this method.

In order to calculate these amplitudes we follow the rules from
helicity formalism and use identities of the type

\begin{eqnarray}
\{\bar u_{\lambda}(p_{1})\gamma
^{\mu}u_{\lambda}(p_{2})\}\gamma_{\mu} &=& 2u_{\lambda}(p_{2})\bar
u_{\lambda}(p_{1})+ \nonumber \\ && 2u_{-\lambda}(p_{1})\bar
u_{-\lambda}(p_{2}),
\end{eqnarray}

\noindent which is in fact the so called Chisholm identity, and

\begin{equation}
p\llap{/}=u_{\lambda}(p)\bar u_{\lambda}(p)+u_{-\lambda}(p)\bar u_{-\lambda}(p),
\end{equation}

\noindent defined as a sum of the two projections $u_{\lambda}(p)\bar u_{\lambda}(p)$
and $u_{-\lambda}(p)\bar u_{-\lambda}(p)$.

The spinor products are given by

\begin{eqnarray}
s(p_{i}, p_{j})&\equiv&\bar u_{+}(p_{i})u_{-}(p_{j})=-s(p_{j}, p_{i}),\nonumber\\
t(p_{i}, p_{j})&\equiv&\bar u_{-}(p_{i})u_{+}(p_{j})=[s(p_{j}, p_{i})]^{*}.
\end{eqnarray}

Using the above rules, which are proved in Ref. \cite{stirling}, we
can reduce many amplitudes to expressions involving only spinor products.

For the polarization of the initial photon we take \cite{stirling}
$\epsilon^{\mu}_{\lambda}(k)=N \bar
u_{\lambda}(k)\gamma^{\mu}u_{\lambda}(p)$, where $N=1/\sqrt{4
k.p}$ and $p^{\mu}$ is any lightlike vector not collinear to
$k^{\mu}$. We take for $p^{\mu}$ one of the other momenta
occurring in the problem. In particular we take $p=p_1$, where
$p_1$ is the 4-moment of the tau lepton.

For simplicity in the expressions and in the numerical calculation
we assign a number for each 4-moment as it is shown in Fig.1. In
this conditions, we represent the products $s(p_i,p_j)$ and
$t(p_i,p_j)$ with the symbols $s_{ij}$ and $t_{ij}$ respectively.
The corresponding amplitude are written as functions of these
symbols:

\begin{eqnarray}
T_1(-1,-1,-1,-1)&=&- 4f_1 f_2 s_{13}s_{34}t_{12}t_{13} \nonumber
\\
 T_1(-1,-1,-1,\mbox{\hspace{3mm}}1)&=&4 f_1 f_2 s_{13}s_{24}t_{13}^2
(-1+4c_{\tau}^2s_{23}t_{23}) \nonumber \\
T_1(-1,\mbox{\hspace{3mm}}1,-1,-1)&=&8 f_1 f_2 c_{\tau}
s_{13}s_{23}t_{12}t_{13}t_{24} \nonumber \\
 T_1(-1,\mbox{\hspace{3mm}}1,-1,\mbox{\hspace{3mm}}1)&=& -
8 f_1 f_2 c_{\tau}  s_{13}s_{23}t_{13}^2 t_{34} \nonumber
\\ T_1(\es1,-1,\es1,-1)&=& -
8  f_1 f_2 c_{\tau} s_{13}^2 s_{34}t_{13}t_{23} \nonumber
\\ T_1(\mbox{\hspace{3mm}}1,-1,\mbox{\hspace{3mm}}1,\mbox{\hspace{3mm}}1)&=&
8 f_1 f_2 c_{\tau}  s_{12} s_{13}s_{24}t_{13}t_{23} \nonumber
\\ T_1(\es1,\es1,\es1,-1)&=& 4 f_1 f_2 s_{13}^2 t_{13} (-1 +
4 c_{\tau}^2 s_{23}t_{23}) t_{24}
 \nonumber \\ T_1(\es1,\es1,\es1,\es1)&=& -
4 f_1 f_2 s_{12}s_{13}t_{13}t_{34}
 \nonumber \\
T_2(-1,-1,-1,-1) &=& 4 f_1 f_2 s_{23} s_{34} t_{12} t_{23}
\nonumber
\\
 T_2(-1,-1,\es1,-1) &=& 4 f_1 f_2 s_{14} s_{23}(-1 +
4 c_{\tau}^2 s_{13} t_{13}) t_{23}^2
 \nonumber \\
T_2(-1,\es1,-1,-1)&=&-8 f_1 f_2 c_{\tau}  s_{13} s_{23} t_{12}
t_{14} t_{23} \nonumber
\\ T_2(-1,\es1,\es1,-1)&=&-8 f_1 f_2 c_{\tau} s_{13} s_{23}
t_{23}^2 t_{34} \nonumber \\ T_2(\es1,-1,-1,\es1)&=&-8 f_1 f_2
c_{\tau}  s_{23}^2 s_{34} t_{13} t_{23} \nonumber
\\ T_2(\es1,-1,\es1,\es1)&=&-8  f_1 f_2 c_{\tau}
s_{12} s_{14} s_{23} t_{13} t_{23} \nonumber
\\
 T_2(\es1,\es1,-1,\es1) &=& 4 f_1 f_2 s_{23}^2 (-1+
4 c_{\tau}^2 s_{13} t_{13}) t_{14} t_{23} \nonumber
\\ T_2(\es1,\es1,\es1,\es1)&=& 4 f_1 f_2 s_{12} s_{23} t_{23}
t_{34}
\end{eqnarray}

where $c_{\tau}=a_{\tau}/(4 m_{\tau})$ and the arguments of the
$T$ functions correspond to the helicities of $\tau$, $\bar\tau$
and the photons respectively.
 After the evaluation of the amplitudes for the
corresponding diagrams,
 we obtain
the cross-sections of the analyzed processes for each point of the
phase space and for different helicity of the particles involved
in the process.

In this conditions we are ready to study different kinds of
observables built with the cross-section for polarized particles.
The photon, which goes ahead in positive z direction, is considered
 polarized left, being z the beam axis. The other initial photon,
that goes ahead in the negative z direction, is supposed unpolarized.
 Thus, we define the following observable and we call it
Asymmetry

\begin{eqnarray}
{\cal A}_{LR}(\theta)=\frac{d\sigma/d\Omega|_L-d\sigma/d\Omega|_R}
{d\sigma/d\Omega|_L+d\sigma/d\Omega|_R}
\end{eqnarray}

where $d\sigma/d\Omega|_{L(R)}$ is the differential
cross-section production for Left (Right) tau. The anti-tau is considered
unpolarized and the initial state is prepared as we have explained
above. From the general expression for the amplitude is easy to
obtain an analytic result for ${\cal A}_{LR}(\theta)$

\begin{eqnarray}
{\cal A}_{LR}(\theta)=\frac{-2
\cos\theta+4c_{\tau}^2s(2-\cos\theta)\sin^2\theta}
{1+\cos^2\theta+16 c^2_{\tau} s (1+c^2_{\tau} s
\sin^2\theta/4)\sin^2\theta}
\end{eqnarray}

In fig.4 and fig.5 we show ${\cal A}_{LR}(\theta)$ for different
values of $a_{\tau}$ for $\sqrt{s}=200$GeV and $500$GeV
respectively. In particular we plot the curve for $a_{\tau}=0.009$
which corresponds to the best bound obtained from other authors
\cite{gabriel}, and the curve for $a_{\tau}=0.001177$ that
corresponds to the theoretical estimation for $a_{\tau}$. As we
can see this asymmetry is an observable sensible to effects of
magnetic anomalous moment. In particular for $\sqrt{s}=500$ GeV,
we can see an important apartment between the standard prediction
($a_{\tau}=0$) and the best present bound ($a_{\tau}=0.009$). In
this respect we define the relative apartment
between the asymmetry and the SM-asymmetry ($a_{\tau}=0$)

\begin{equation}
\delta=\left|\frac{{\cal A}_{LR}(\theta)-{\cal
A}^{SM}_{LR}(\theta)}{2}\right|\times 100
\end{equation}
where the denominator $2$ in the above expression corresponds to
the all variation range of the asymmetry. This apartment $\delta$
can be think as related to the experimental precision in the
${\cal A}_{LR}$ measurement. In fig. 6 and 7 we plot $\delta$ for
$\sqrt{s}=200$GeV and $\sqrt{s}=500$GeV respectively. We can see
a significant apartment for moderate values of $a_{\tau}$, which
makes of this asymmetry a useful observable to bound magnetic
anomalous moment.

For completeness, we define another related observable involving
the total cross-section

\begin{eqnarray}
 B_{LR}=\frac{\sigma_L-\sigma_R} {\sigma_L+\sigma_R}
\end{eqnarray}

where

\begin{equation}
\sigma_{L(R)}=\int \frac{d\sigma}{d\Omega}|_{L(R)} d\Omega
\nonumber
\end{equation}

For this observable, it is easy to obtain an analytic expression
\begin{equation}
B_{LR}=\frac{c_{\tau}^2 s} {1/4+2 c_{\tau}^2 s+2/5 c_{\tau}^4 s^2
}
\end{equation}

Note that this observable vanishes in the Standard Model
($a_{\tau}=0$). In fig.8 we show it as a function of $a_{\tau}$
for $200$ GeV and $500$ GeV center of masses energies
respectively. We include a vertical line that represent the current best
bound for $a_{\tau}$. As we can see the $B_{LR}$ values show
a significant deviation of the SM value ($B_{LR}=0$).

\section{Conclution}

In this work we study a set of different observables which were
not investigated by another authors previously. These observables
involve universality test and polarization effects. We have
founded that, by using these observable set, it could be possible to
improve the actual bounds for $a_{\tau}$.

{\bf Acknowledgements}

We thank CONICET (Argentina), Universidad Nacional de Mar del
Plata (Argentina) for their financial supports.

\pagebreak

\noindent{\large \bf Figure Captions}\\

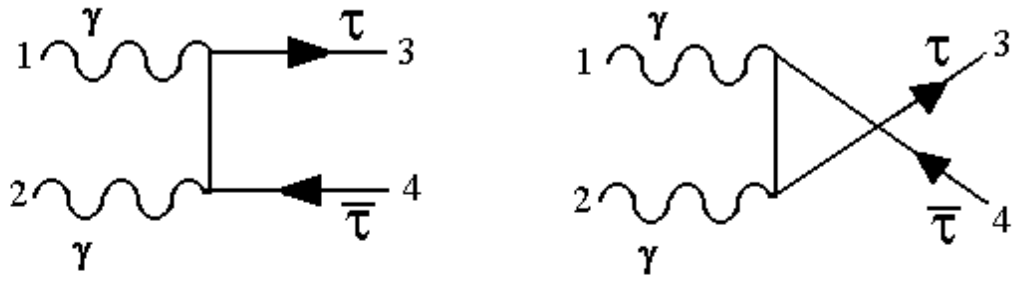
\noindent{\bf Figure 1:} Feynman graph contributing to the
amplitude of the $\gamma \gamma \rightarrow \tau \bar{\tau}$
process.

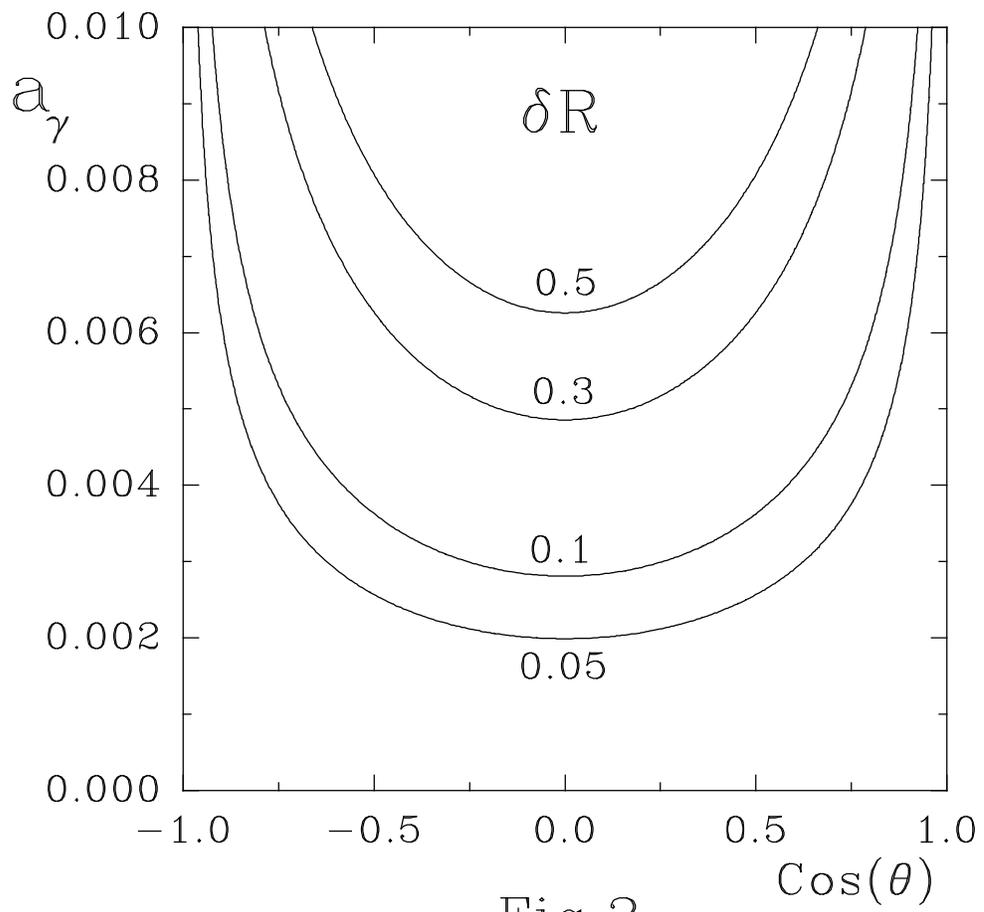
\noindent{\bf Figure 2:} Contour plot for $\delta R$ in the
($\cos\theta,a_{\tau}$) plane for $\sqrt{s}=200$GeV.

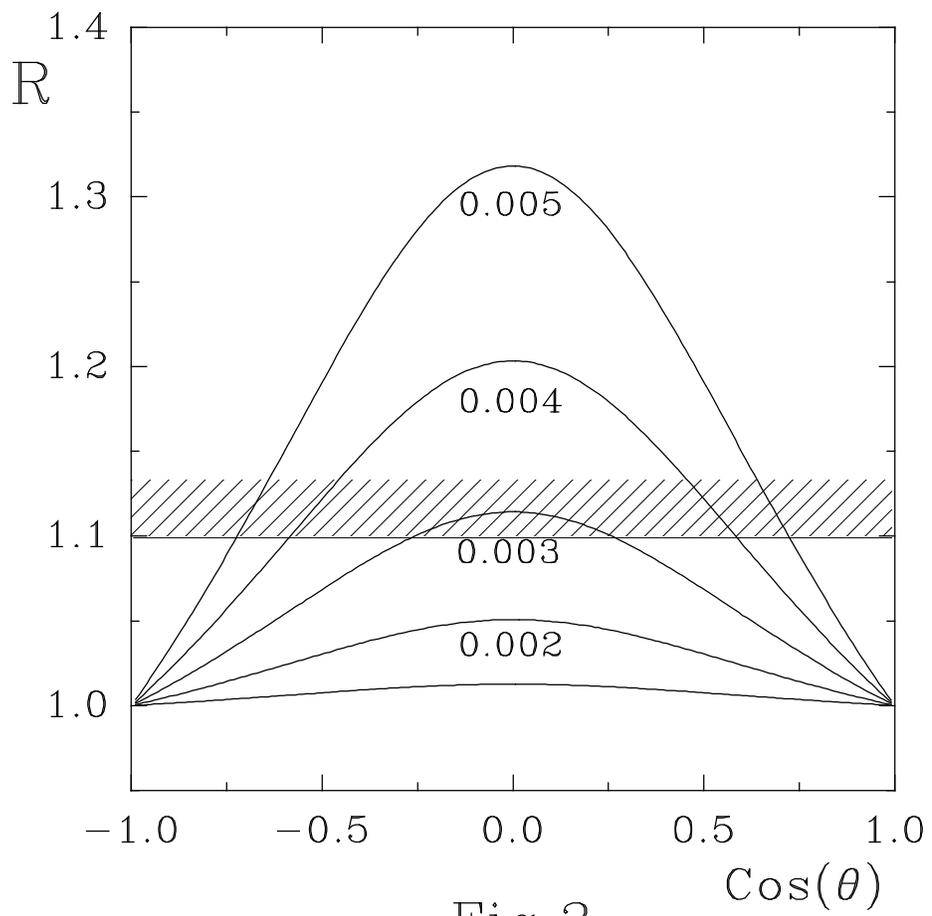
\noindent{\bf Figure 3:}Universality ratio for $\sqrt{s}=200$GeV
and for different values of $a_{\tau}$. We have included the limit
for an hypothetical measurement of $R$ with a accuracy of $10\%$.

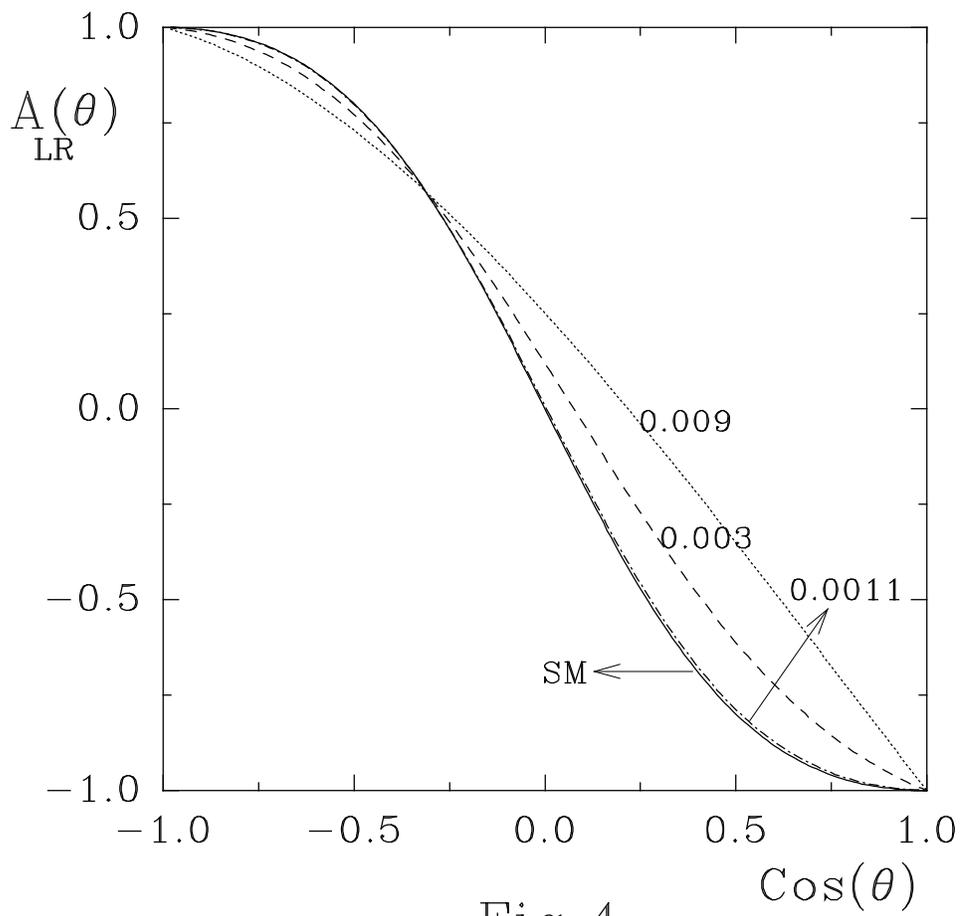
\noindent{\bf Figure 4:} The angular dependent asymmetry ${\cal
A}_{LR}(\theta)$ as a function of $\cos\theta$ for
$\sqrt{s}=200$GeV and for different value of $a_{\tau}$. We
include the curves corresponding to the best bound for
$a_{\tau}$ and for the theoretical estimation.

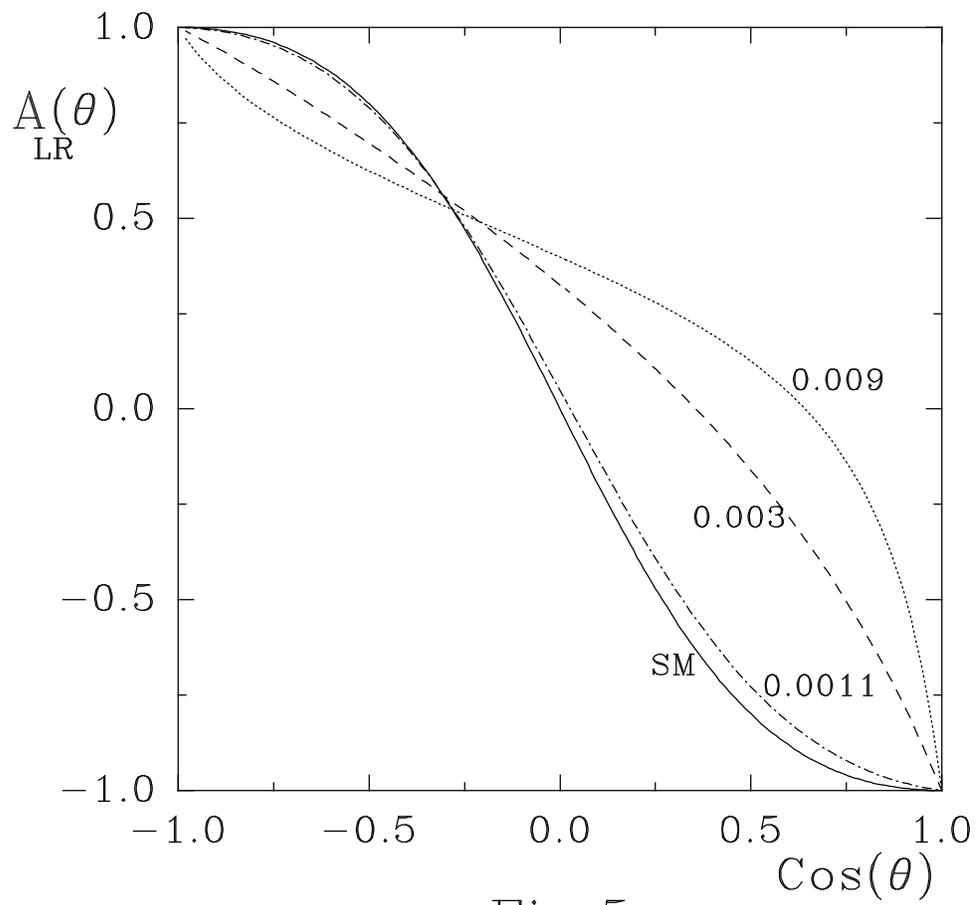
\noindent{\bf Figure 5:} The same of fig.4 but for $\sqrt{s}=500$
GeV.

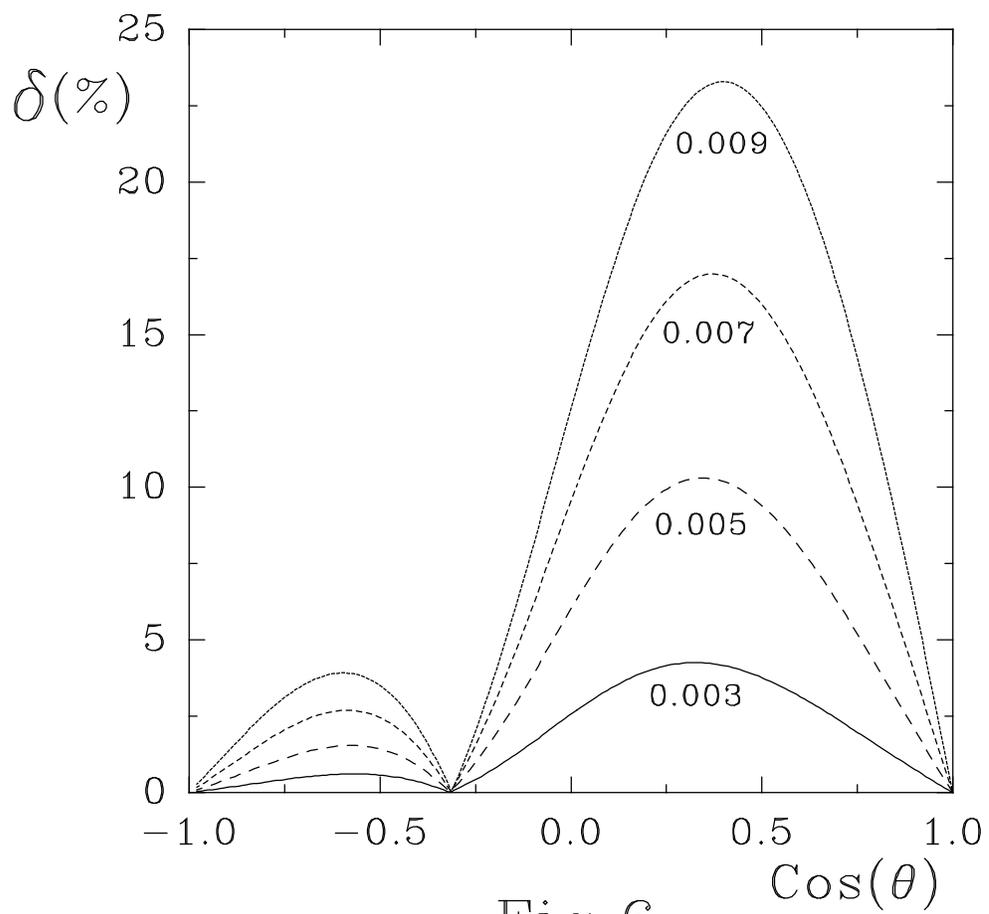
\noindent{\bf Figure 6:} $\delta$ as a function of $\cos\theta$
for different value of $a_{\tau}$ for $\sqrt{s}=200$ GeV.

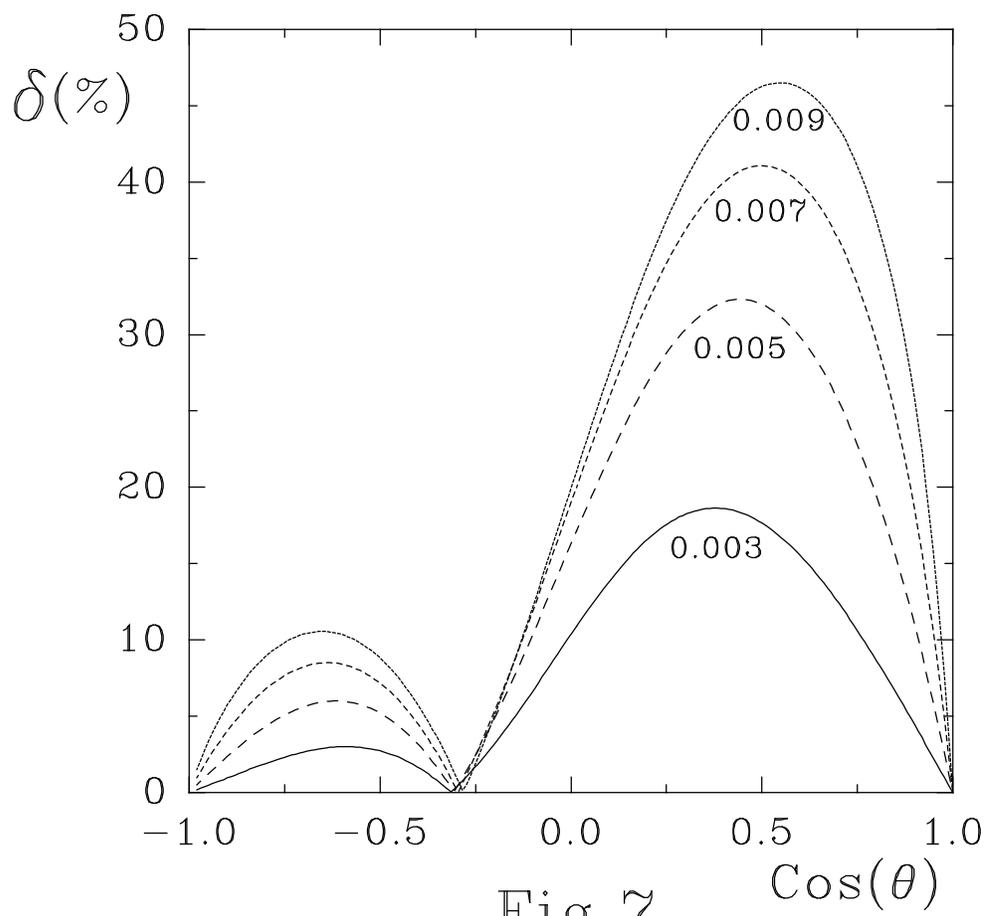
\noindent{\bf Figure 7:} The same of fig.6 but for $\sqrt{s}=500$
GeV.

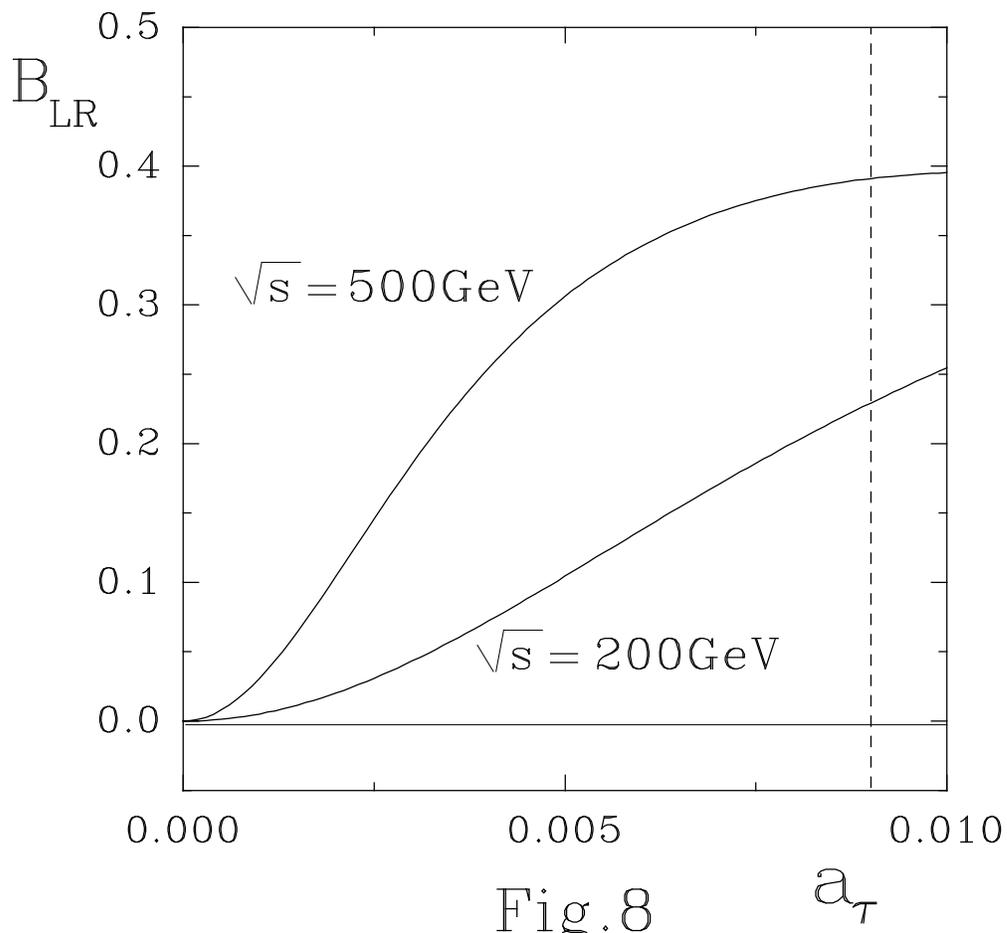
\noindent{\bf Figure 8:} The integrate asymmetry $B_{LR}$ as a
function of $a_{\tau}$ for different values of center mass
energies. The vertical line represent the present best bound
obtained for $a_{\tau}$.

\end{document}